\documentclass{PoS}

\newcommand{\AmS}{{\protect\the\textfont2
  A\kern-.1667em\lower.5ex\hbox{M}\kern-.125emS}}

  \newcommand \beq{\begin{eqnarray}}
\newcommand \eeq{\end{eqnarray}}
\newcommand{\bea}{\begin{eqnarray}}
\newcommand{\eea}{\end{eqnarray}}
\newcommand{\nn}{\nonumber\\ }

\def\simle{\mathrel{\rlap{\raise 0.511ex \hbox{$<$}}{\lower 0.511ex \hbox{$\sim$}}}}
\def\simge{\mathrel{ \rlap{\raise 0.511ex \hbox{$>$}}{\lower 0.511ex \hbox{$\sim$}}}}

\def\del{\partial}

\def\Im{{\,\rm Im\,}}
\def\Re{{\,\rm Re\,}}
\def\tr{{\,\rm tr\,}}
\def\0{\over } \def\2{{1\over2}} \def\4{{1\over4}}
\def\5{\hat } \def\6{\partial }

\def\8#1{{\textstyle{#1}}}

\def\({\left(} \def\){\right)} \def\<{\langle } \def\>{\rangle }

\title{Exact renormalization group at finite temperature}

\ShortTitle{RG at finite T}

\author{\speaker{JeanPaul Blaizot}\thanks{member of CNRS}\\
        IPhT-CEA Saclay 91191 Gif-sur-Yvette cedex, France\\
        E-mail: \email{jean-paul.blaizot@cea.fr}}

\abstract{This talk  reviews progress in the (semi-) analytic calculations of the thermodynamics of the quark-gluon plasma. I shall explain how weak coupling techniques can allow us, through appropriate resummations, to deal with particular non perturbative aspects of the quark-gluon plasma. Then I shall show how the exact renormalization group may provide insight into the physics of this multi-scale system.\
}

\FullConference{International Workshop on QCD Green's Functions, Confinement, and Phenomenology - QCD-TNT09\\
		 September 07 - 11 2009\\
		 ECT Trento, Italy}

\begin{document}

\section{Introduction}

There is presently much discussion going on in order to understand the origin of the strongly coupled character of the quark-gluon plasma revealed by the RHIC  data on heavy ion collisions \cite{RHIC}. These data have indeed led to a paradoxical situation. 

On the one hand, as a natural consequence of 
asymptotic freedom, one expects hadronic matter  to turn, at high
temperature and density, into a gas of quarks and gluons whose free
motion is only weakly perturbed by their interactions. Weak coupling calculations (based on resummed QCD perturbation theory) reproduce lattice results for temperatures greater than 2.5 to 3 $T_c$  \cite{Blaizot:2000fc}, and provide support to a quasiparticle picture: according to these calculations, the dominant effect of the interactions is to turn (massless) quarks and gluons into weakly interacting (massive) quasiparticles.  
Recent lattice calculations, that can probe arbitrarily large temperatures,  demonstrate the approach to the Stefan-Boltzmann  limit in a convincing way, in good agreement with weak coupling calculations \cite{Endrodi:2007tq}. 
The calculations of the fluctuations of conserved charges (such as baryon number, electric charge, strangeness) provide another evidence that the bulk quark behavior resembles that  a free gas  above the deconfinement transition \cite{Cheng:2008zh}. 

On the other hand, RHIC data do not provide any evidence for ideal gas
behavior \cite{RHIC}. Indeed, the strong opacity of matter to hard parton propagation (see e.g.  \cite{Majumder:2007iu}
 for a recent discussion), the strong collective elliptic flow, and the low value of the ratio $\eta/s$ (where $\eta$ is the shear viscosity and $s$ the entropy density) \cite{Luzum:2008cw}, are observations that picture the quark-gluon plasma produced at RHIC as a strongly coupled system. 

One may argue that,  at RHIC, the quark-gluon plasma spends most of its existence in a region, say between  $T_c$ and  $\sim  3T_c$, where the physics is hard and poorly understood. It seems indeed that, in this region, the quasiparticle picture breaks down, and genuine non-perturbative effects appear in bulk thermodynamics.  For instance $\epsilon-3P$,  the trace of the energy momentum tensor is non vanishing, reflecting the QCD scale anomaly.  Still, even in that region, explicit calculations reveal that the coupling constant is not huge (near $T_c$, $g\sim 2$, see e.g. \cite{Laine:2005ai}).

At this point, it is worth recalling that non perturbative features may arise in a system from the 
cooperation of many degrees of freedom, or 
strong classical fields, making the system  strongly interacting while the elementary coupling strength remains small. In the quark-gluon plasma coexist degrees of freedom with different wavelengths. Whether these degrees of freedom are weakly or strongly coupled depends crucially on their wavelength: short wavelength degrees of freedom may be weakly coupled if the coupling constant is small, while very long wavelength modes will remain strongly coupled, however small the coupling constant may be. 

These remarks will help us to understand that, while strict perturbation theory breaks down even at moderate values of the coupling constant, weak coupling techniques with appropriate resummations can lead to accurate calculations of thermodynamical properties of the quark-gluon plasma. This is because short wavelength degrees of freedom dominate the thermodynamics: the long wavelength modes carry individually little energy and they have limited phase space, so their total contribution to the pressure is small. Short wavelength degrees of freedom can be handled with a combination of effective field theory and perturbation theory, as we shall see in the next section.

\section{Weak coupling techniques}

\subsection{Breakdown of strict perturbation theory}

 Much effort has been put into calculating
 the successive orders of the perturbative expansion for the pressure
and
the series is known now up to order $g^6\ln
g$ (see \cite{Kajantie:2002wa} and references therein).  These calculations have revealed  that
perturbation theory  makes sense only for very small values of the
coupling constant, corresponding to extremely large values of $T$.
For not too small values of the coupling, the successive terms in
the expansion oscillate wildly and the dependence of the results on
the renormalization scale keeps increasing order after order (see e.g.
\cite{Blaizot:2003tw}), making strict perturbation theory inapplicable to estimate the corrections to the ideal quark
gluon plasma. Let us emphasize that the problem is not specific to QCD, but also occurs in simpler scalar field theories (see  \cite{Blaizot:2003tw}, and also Fig.~\ref{fig1} below). 

 This situation is to be contrasted with what happens at zero temperature,
 where perturbative calculations achieve a reasonable accuracy already at the GeV scale.
 The point is that, as we already mentioned,  the validity of the weak coupling expansion depends
not only on the strength of the coupling, but also on the number of
active degrees of freedom. At zero temperature, one deals most of
the time with a very limited number of degrees of freedom (the
colliding particles and the reaction products), while at finite
temperatures, as we shall see shortly, the  thermal fluctuations
alter the infrared behavior in a profound way.

\subsection{The role of thermal fluctuations}
In the quark-gluon plasma, the effect of the interactions at a given scale depends
on the magnitude of the relevant thermal fluctuations at that scale (and in some cases at a different scale as well).  Let us  examine this in detail. In order to avoid being distracted by issues related to gauge invariance we shall restrict ourselves to the case of a simple massless scalar field theory, with a $g^2\phi^4$ interaction. The thermal fluctuations are given by the following integral
\beq\label{fluctuations}
\langle \phi^2\rangle= \int\frac{d^3 k}{(2\pi)^3}\frac{n_k}{k}, \qquad n_k=\frac{1}{{\rm e}^{k/T}-1}.
\eeq
When we perform a perturbative calculation, we assume that the ``kinetic energy'' $\sim \langle ( \del\phi)^2\rangle$ is large compared to the ``potential energy'' $\sim g^2 \langle\phi^4\rangle$. Obviously, this comparison depends on the strength of the coupling, but also on the typical wavelength, or momentum, of the fluctuations. To make things more precise, let us observe that the integral (\ref{fluctuations}) is dominated by the largest values of $k$ (in the absence of the statistical factor it would be quadratically divergent). One may then calculate the integral with an upper cut-off $\kappa$ and refer to the corresponding value as to ``the contribution of the fluctuations at scale $\kappa$'', and denote it by $\langle \phi^2\rangle_\kappa$. In the same spirit, we shall approximate the kinetic energy as $ \langle ( \del\phi)^2\rangle_\kappa \approx \kappa^2 \langle \phi^2\rangle_\kappa$. Taking furthermore $ \langle\phi^4\rangle_\kappa \approx \langle \phi^2\rangle_\kappa^2$, one gets as expansion parameter
\beq
\gamma_\kappa=\frac{g^2 \langle \phi^2\rangle_\kappa}{\kappa^2}.
\eeq

Let us then examine this parameter for several characteristics momenta. The fluctuations that dominate the energy density at weak coupling correspond to the plasma particles and have momenta $k\sim T$. For these ``hard'' fluctuations, 
\beq\label{fluctuationsT}
\kappa\sim T,\qquad  \langle \phi^2\rangle_T\sim T^2,\qquad \gamma_T\sim g^2.
\eeq
Thus, at this scale,  perturbation
theory works as well as at zero temperature (with expansion parameter $\sim g^2$, or rather $\alpha=g^2/4\pi$). 

The next ``natural'' scale, commonly referred to as the ``soft scale'', corresponds to $\kappa\sim gT$. We have
\beq\label{fluctuationsgT}
\kappa\sim gT,\qquad \langle \phi^2\rangle_{gT}\sim gT^2,\qquad \gamma_{gT}\sim g.
\eeq
In calculating $\langle \phi^2\rangle_\kappa$ for $\kappa\ll T$, we have used the approximation $n_k\approx T/k$, so that $\langle \phi^2\rangle_{\kappa\ll T}\sim \kappa T$. We note that $ \gamma_{gT}\sim g$, so that perturbation theory can still be used to describe the self-interactions of the soft modes. However the perturbation theory is now an expansion in powers of $g$ rather than $g^2$: it is therefore less rapidly convergent. The emergence of this new expansion parameter is the origin of odd powers of $g$ in the perturbative expansion of the pressure  (such as the plasmon term $\sim g^3$). 

Another phenomenon occurs at the scale $gT$. While the expansion parameter $ \gamma_{gT}$ that controls the self-interactions of the soft fluctuations is small, the coupling between the soft modes and thermal fluctuations at scale $T$ is not: indeed $g^2\langle \phi^2\rangle_T\sim (gT)^2$. Thus the dynamics of soft modes is non-perturbatively renormalized by their coupling to hard modes. This particular coupling is encompassed by the so-called hard thermal loops \cite{HTL}. 

Finally, there is yet another scale, the ``ultra-soft scale'' $\kappa\sim  g^2 T$, at which perturbation theory completely breaks down. At this scale, we have indeed
\beq\label{fluctuationsg2T}
\kappa\sim g^2T,\qquad \langle \phi^2\rangle_{gT}\sim g^2T^2,\qquad \gamma_{g^2T}\sim 1.
\eeq
Thus the ultra-soft  fluctuations remain strongly coupled for arbitrarily small couplings. Of course, this situation does not really occur for a scalar field since a mass is generated at scale $gT$ which renders the contribution of the fluctuations at the scale $g^2T$ negligible.  However this situation is met in QCD for the long wavelength, unscreened, magnetic
fluctuations. 

These considerations suggest that  the  main difficulty with thermal perturbation
theory is 
not so much related to the fact that the coupling is not small
enough (for the relevant temperatures it is not huge, as we have already pointed out), but
rather to the interplay of degrees of freedom with various
wavelengths, possibly involving collective modes.  Our main concern here is bulk thermodynamics,  dominated by hard degrees of freedom. As we shall see in the next subsections, this can be handled adequately by weak coupling techniques  involving  appropriate reorganizations
and resummations of the perturbative expansion. We shall consider two examples of such approaches. 

\subsection{Effective theory- dimensional reduction}

 A powerful technique to handle situations where modes at different scales couple is that of effective theories. In the present context, it is natural to isolate the  mode carrying zero Matsubara frequencies, which leads to the so-called dimensional reduction. In its more elaborate version \cite{Braaten:1995jr}, this consists in writing a sequence of effective theories, obtained by integrating successively the fluctuations at scale $2\pi T$ and  $gT$. The coefficients of the resulting effective lagrangians can be determined perturbatively as a function of the gauge coupling $g$. 
 
The QCD partition function can be written as ($V$ is the volume) 
 \beq
 P_{QCD}=\frac{T}{V}\ln Z_{QCD}, \qquad  Z_{QCD}=\int DA_k^a\, DA_0^a\, {\rm e}^{-\int d^4 x \,{\cal L}_{QCD}},
 \eeq
 where, to simplify the writing, we ignore the gauge fixing terms. By integrating out the modes with momenta $\sim 2\pi T$, one can re-write this as
 \beq
 P_{QCD}= P_E+\frac{T}{V}\ln Z_{E} 
 \eeq
 where the lagrangian ${\cal L}_E$ in $Z_E$ is is that of the effective theory at scale $gT$, and is of the form
 \beq
 {\cal L}_E=\frac{1}{2}{\rm Tr}F_{kl}^2+{\rm Tr}[D_k,A_0]^2+m_E^2{\rm Tr} A_0^2+\lambda_E \left({\rm Tr} A_0^2  \right)^2+\cdots
 \eeq
 where the dots denote higher dimension operators. Here the fields depend only on the spatial coordinates, $D_k=\del_k-ig_E A_k$, and the various parameters are determined using perturbation theory. In leading order, $m_E^2\sim g^2 T^2$, $g_E^2\sim g^2T$, $\lambda_E\sim g^4T$, while $P_E\sim T^4$. 
 
 Integrating the modes at scale $gT$ leaves an effective theory for the modes at scale $g^2T$:
 \beq
\frac{T}{V}\ln Z_E=p_M+\frac{T}{V} \ln Z_M,\qquad  Z_M=\int DA_k^a \; {\rm e}^{-\int d^3 x\, {\cal L}_{M}}
, \qquad {\cal L}_M=\frac{1}{2} {\rm Tr}F_{kl}^2+\cdots
 \eeq
 In leading order, $p_M\sim m_E^3T$, $g_M^2\sim g_E^2$. This effective theory is equivalent in its leading order  to a 3 dimensional Yang-Mills theory. It gives  a non  perturbative contribution to the pressure of order $g^6$.

 Calculations based on this scheme have been pushed to high order \cite{Kajantie:2002wa}, but the determination of the order $g^6$ contribution to the pressure  depends on an as yet undetermined 4-loop matching
coefficient. By adding a parameter to account for this uncalculated contribution, one can match the four-dimensional lattice results in a range of temperatures between $2.5 T_c$ and $3T_c$.  It is interesting to note that  better results are obtained (in particular a weaker dependence on the renormalization scale) if one refrains from expanding the parameters of the effective theory in terms of the original gauge coupling  $g$ \cite{Blaizot:2003iq}.

\subsection{Skeleton expansion- 2PI approximations}

\begin{figure}
\begin{center}
\vspace{-70mm}
\includegraphics[scale=0.6]{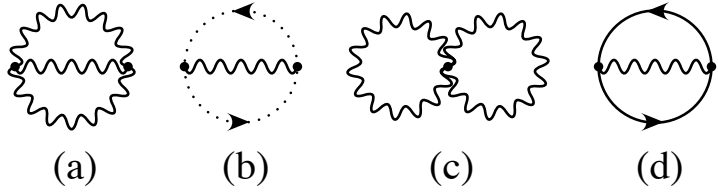}
\vspace{-60mm}
\caption{Skeleton (2PI) diagrams that are taken  into account at 2-loop order in QCD for the calculation of the entropy according to Eq.~(2.10).
Wiggly, plain, and dotted lines refer respectively to
gluons, quarks, and ghosts.
  \label{figphiqcd} }
\end{center}
\end{figure}

\begin{figure}
\begin{center}
\vspace{-50mm}
\includegraphics[scale=0.5]{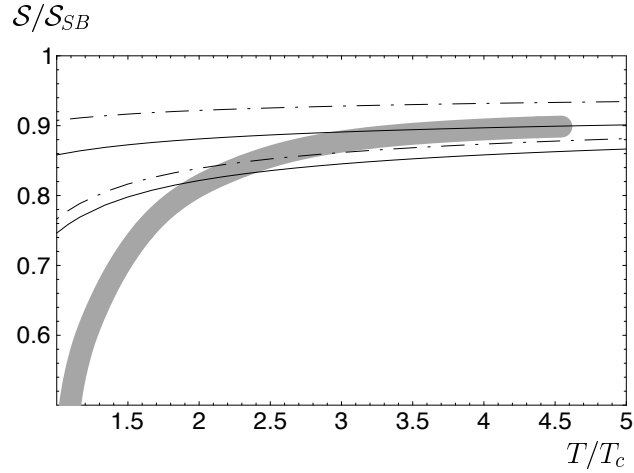}
\vspace{-30mm}
\caption{Comparison of the lattice data (grey band) for the entropy of pure gauge SU(3) theory \cite{Boyd:1996bx} with two successive approximations (solide and dashed lines) in the two-loop 2PI formalism  (from \cite{Blaizot:1999ip}).}
\label{figS}
\end{center}
\end{figure}

There exists other ways to reorganize the perturbative expansion (for a review see \cite{Blaizot:2003tw}).
I shall briefly recall here an approach based on a $\Phi$-derivable
two-loop approximation (also called  2PI formalism) applied to the calculation of the
entropy density  \cite{Blaizot:1999ip,Blaizot:2000fc}:
\beq
\label{S2loop}
{\cal S}&=&-\tr \int{d^4k\0(2\pi)^4}{\6n(\omega)\0\6T} \left[ \Im
\log D^{-1}-\Im \Pi \Re D \right] \nn
&&-2\tr \int{d^4k\0(2\pi)^4}{\6f(\omega)\0\6T} \left[ \Im
\log S^{-1}-\Im \Sigma \Re S \right],
\eeq
with the relevant Feynman diagrams displayed in Fig.~\ref{figphiqcd}.
In contrast to dimensional reduction, based on the imaginary time formalism,  this approach exploits the real-time information that one has on the elementary excitations of the quark-gluon plasma: the spectral information enters the calculation of the entropy through the propagators of gluons ($D$) and quarks ($S$), and the respective self-energies ($\Pi$ and $\Sigma$). It was shown in  
Refs.~\cite{Blaizot:1999ip,Blaizot:2000fc} that the lattice results for
the entropy of the gluonic  plasma  were quite well
reproduced  for $T \ge 3 T_c$ (see Fig.~\ref{figS}). 

The  formalism used in this calculation of the entropy    has been tested \cite{Blaizot:2005wr} in the limit of a large number of quark flavors, which can be solved exactly. One then found that the 2PI approximation scheme allows for a smooth extrapolation that is accurate up to quite large values of the coupling constant.

This approach has also the virtue of providing a clear physical picture:  at high $T$, the dominant effect of the interactions is to turn  the original (massless) degrees of freedom, quarks and gluons, into massive quasiparticles, with weak residual interactions.   As shown by  Fig.~\ref{figS}, this picture is consistent with lattice calculations
for temperatures above $ 3T_c$. Note that in the region below $2.5T_c$ both this approximation and that based on dimensional reduction systematically overshoot the lattice data, suggesting that different  physics is involved there.

\subsection{Strong coupling and Supersymmetric Yang Mills theories}

A radically different approach consists in assuming that all the degrees of freedom of the quark-gluon plasma are strongly coupled, and take this strongly coupled quark-gluon plasma as a starting point for further approximations (such as a strong coupling expansion). This is made possible by a theoretical breakthrough that allows calculations in (some) strongly coupled gauge theories, based on the so-called AdS/CFT  duality (for a recent review, see e.g. \cite{Gubser:2009fc}). One  prediction of such calculations concerns the entropy $S$  of ${\cal N}=4$  supersymmetric Yang-Mills (SYM) theory which behaves, in strong coupling, as \cite{Gubser:1998nz}
\beq
\frac{S}{S_0}=\frac{3}{4}+\frac{45}{32}\zeta(3)\frac{1}{\lambda^{3/2}},
\eeq 
where $\lambda\equiv 2g^2 N_c$ and $S_0$ is the entropy of the non interacting system. Thus,
 in the limit of strong coupling, $\lambda\to\infty$,
the entropy is bounded from below by the value $S/S_0=3/4$, a value that is not too distinct from that obtained from QCD lattice calculations for temperatures above $3T_c$. In fact, in this temperature range, the entropy
density is about half-way between its weak coupling value and its
strong coupling value, and there seems to be  no compelling reason to favor  an interpretation of lattice data based  on strong coupling (see
\cite{Blaizot:2006tk} for a recent discussion).

\section{Insights from the exact renormalization group}

As the preceding  discussion has shown, the quark-gluon plasma is a multi-scale system. When the coupling is small, a clean hierarchy of well separated momentum scales emerges ($T, gT, g^2T$). This allows us to treat the very high temperature plasma using  a combination of effective theories and perturbation theory. However, as the temperature decreases, the coupling grows and the various scales start to mix, making  the accuracy of such calculations worse and worse. In this last part of the talk, I shall indicate how the non perturbative (sometimes called exact) renormalization group (for a review, see \cite {Berges:2000ew}) can be used to handle this situation.  The present discussion is limited to the case of the scalar field theory: as we have seen, the difficulties of perturbative calculations in QCD have much in common with the corresponding calculations for the scalar field; furthermore we have for this scalar case results corresponding to rather elaborate solutions of the flow equations. (For a recent discussion of the application of the exact renormalization group to QCD see e.g. \cite{Gies:2006wv}.)

\subsection{The exact renormalization group}

\begin{figure}
\vspace{-10mm}
\begin{center}
\includegraphics[scale=0.50]{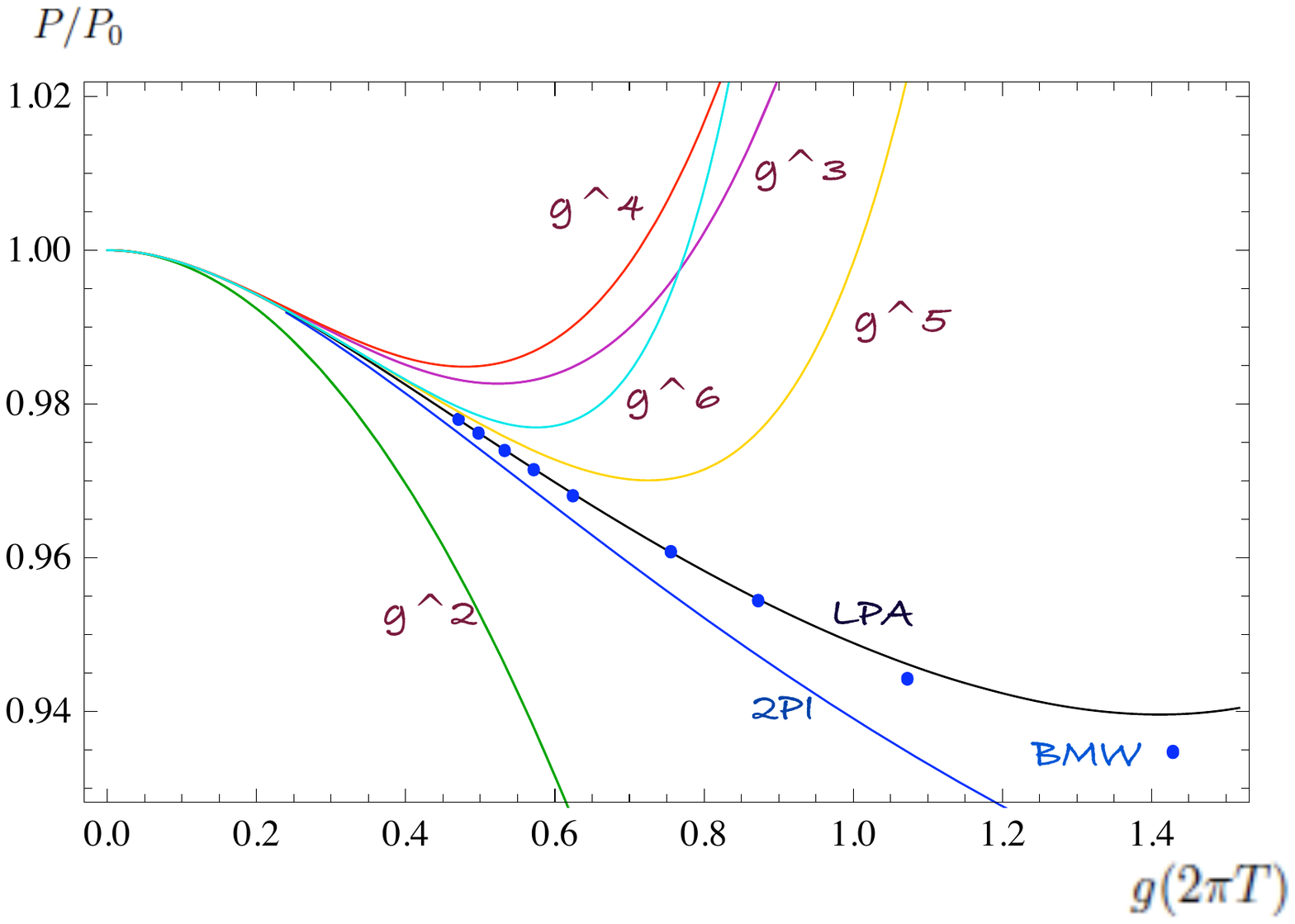}
\vspace{-20mm}
\caption{The pressure as a function of the coupling constant}
\label{fig1}
\end{center}
\end{figure}

 There is
some analogy between the effective field theory approach, such as that mentioned earlier,  and the non
perturbative renormalization group: in effective field theory one
integrates out degrees of freedom above some cut-off; in the
renormalization group this integration is done smoothly.   In a way,
the renormalization group builds up a continuous tower of effective
theories that lie infinitesimally close to each other, and  are
related by a renormalization group flow equation.  This picture is
independent of the value of the coupling, so that the
renormalization group provides  a smooth extrapolation from the
regime of weak coupling, characterized by a clean separation of
scales, towards the strong coupling regime where all scales get
mixed.  In a recent study \cite{Blaizot:2006rj}, it has been shown in the case
of a scalar  $\phi^{4}$ theory   with $O(N)$ symmetry, that such a
technique can provide  a smooth extrapolation to strong coupling,
which turns out to be similar to  that of a simple (2-loop) 2PI
approximation.

In practice, the exact renormalization group   builds a family of theories indexed by a 
momentum scale parameter $\kappa$,   
such that fluctuations are smoothly taken into account  as $\kappa$ is lowered 
from the  microscopic scale $\Lambda$ down  to 0. 
This is achieved by adding  to the original  
Euclidean action $S$ a  term of the form   
$\Delta S_\kappa[\varphi]= \frac{1}{2} \int_q\: R_\kappa(q)\varphi(q)\varphi(-q)$, where the  cut-off function $R_\kappa(q)$ is chosen so that: i) $R_\kappa(q)\sim \kappa^2$ for $q\lesssim \kappa$, which 
effectively suppresses the modes $\varphi(q\lesssim  \kappa)$, and ii) 
$R_\kappa(q)$ vanishes for $q\gtrsim \kappa$, leaving 
the modes $\varphi(q\gtrsim \kappa)$ unaffected. 
 One can write for
$\Gamma_\kappa[\phi ]$ an exact flow equation \cite{Wetterich:1992yh}:
 \beq \label{NPRGeq}
\partial_\kappa \Gamma_\kappa[\phi]=\frac{1}{2} \int \frac{d^dq}{(2\pi)^d}
\,\partial_\kappa R_\kappa(q)\,
G_\kappa(q,-q;\phi),
\eeq
where $G_\kappa[\phi]$ is the full propagator in the presence of the background field $\phi$:
\beq
G_\kappa^{-1}[\phi]=\Gamma_\kappa^{(2)}[\phi]+R_\kappa,
\eeq
with $\Gamma^{(2)}_\kappa[\phi]$  the second functional derivative of 
$\Gamma_\kappa[\phi]$ w.r.t. $\phi$. 
The initial conditions of the flow equation (\ref{NPRGeq}) are specified at 
the microscopic scale $\kappa=\Lambda$ where fluctuations are frozen by  
$\Delta S_\kappa$, so that $\Gamma_{\kappa=\Lambda}[\phi]\approx S[\phi]$. 
The effective action of the original scalar field theory  is obtained as the solution of  (\ref{NPRGeq})
for $\kappa\to 0$ where $R_\kappa(q^2)$ vanishes.

Differentiating Eq.~(\ref{NPRGeq}) $m$ times with respect to $\phi$
yields the flow equation for the $m$-point function 
$\Gamma^{(m)}_k[q_1,\dots,q_m;\phi]$. Thus for instance,  the flow equation for $\Gamma^{(2)}$ reads:
\begin{eqnarray}
\label{gamma2champnonnul}
\partial_\kappa\Gamma_{\kappa}^{(2)}(p,\phi)&=&\int
\frac{d^dq}{(2\pi)^d}\,\partial_\kappa R_\kappa(q)\,G_{\kappa}^2(q,\phi)\nonumber\\
&\times&\left\{\Gamma_{\kappa}^{(3)}(p,q,-p-q;\phi) G_{\kappa}(q+p,\phi)\Gamma_{\kappa}^{(3)}(-p,p+q,-q;\phi)
-\frac{1}{2}\Gamma_{\kappa}^{(4)}
(p,-p,q,-q;\phi)\right\} .\nonumber \\\eeq
Note that the flow equation for $\Gamma^{(m)}_k[q_1,\dots,q_m;\phi]$ involves 
$\Gamma^{(m+1)}_k$ and $\Gamma^{(m+2)}_k$, leading to an infinite hierarchy. Approximations need to be done to close this hierarchy. We shall present results obtained with two distinct approximations. 

\subsection{Approximations}

The simplest approximation is the so-called local potential approximation (LPA). It leads to a closed  flow equation for  the effective potential $V_\kappa$. This follows from that of the effective action $\Gamma_\kappa$, Eq.~(\ref{NPRGeq}),  when restricted to constant $\phi$. It reads  \beq\label{eqforV} \kappa\partial_\kappa
V_\kappa(\phi)=\frac{1}{2}\int \frac{d^dq}{(2\pi)^d} \kappa
\,\partial_\kappa R_\kappa(q)\, G_\kappa(q,\phi), \eeq
where 
\begin{equation}\label{G-gamma2}
G^{-1}_{\kappa} (q,\phi) \equiv \Gamma^{(2)}_{\kappa} (q,\phi) +
R_\kappa(q).
\end{equation}
In the local potential approximation one assumes that $\Gamma^{(2)}_{\kappa} (q,\phi) =q^2+\del^2 V/\del\phi^2$, so that,  with this ansatz, the equation for $V$ is indeed a closed equation. 

A more refined approximation \cite{Blaizot:2005xy}  (BMW), consists in neglecting the $q$-dependence of the $m$-point function in the right hand side of the flow equations,
while keeping the full dependence on the external momenta $p_i$.  For instance, in Eq.~(\ref{gamma2champnonnul}), the approximation amounts to the replacements $ \Gamma_{\kappa}^{(3)}(-p,p+q,-q;\phi)\longrightarrow \Gamma_{\kappa}^{(3)}(-p,p,0;\phi)$, and $\Gamma_{\kappa}^{(4)}
(p,-p,q,-q;\phi)\longrightarrow \Gamma_{\kappa}^{(4)}
(p,-p,0,0;\phi)$.
The hierarchy is then closed by observing that 
$\Gamma^{(m+1)}_k(p_1,\dots,p_m,0,\phi)=
\partial_\phi \Gamma^{(m)}_k(p_1,\dots,p_m,\phi)$. 
This approximation has been recently tested on the critical O$(N)$ model and it was found to yield excellent values for the critical exponents \cite{Benitez:2009xg}.

\subsection{Results}

We now present briefly results obtained for the pressure of a massless scalar field theory with a $g^2\phi^4$ coupling \cite{BIW2010}.

Fig.~\ref{fig1} displays the pressure as the function of the coupling constant at the scale $g(2\pi T)$. The various diverging curves indicate the results of perturbative calculations, up to order $g^6$ (for recent high order calculations of the thermodynamics of the scalar field see \cite{Gynther:2007bw,Andersen:2009ct}).
The blue line indicate the result of a 2PI calculation based on a 2-loop skeleton. The black curve indicate the result of the LPA with a Litim regulator, while the blue dots represents the results of the full BMW calculation with an exponential regulator. In contrast to the perturbative calculation, the two calculations based on the renormalization group show a remarkable stability, and a smooth extrapolation towards strong coupling. 

This stability is even more obvious on the next plot, Fig.~\ref{fig2}. Now the pressure is plotted as a function of the thermal mass. This has the advantage of eliminating all scheme dependence \cite{Blaizot:2006rj} (scheme dependence is present when the pressure is plotted as a function of the coupling constant). Note the excellent agreement between the 2PI calculation and the BMW one. Note also the agreement between the LPA and BMW up to large values of the coupling constant. This plot was prepared with  the thermal mass calculated up to order $g^4$, and accordingly the perturbative results for the pressure are reported here only up to this order. 

The stability of these results as one moves towards larger values of the coupling constant, in marked contrast to those of strict perturbation theory, suggests that further corrections to the present calculations (within the renormalization group or the 2PI formalism)  may indeed be very small. 

\begin{figure}
\begin{center}
\vspace{-10mm}
\includegraphics[scale=0.5]{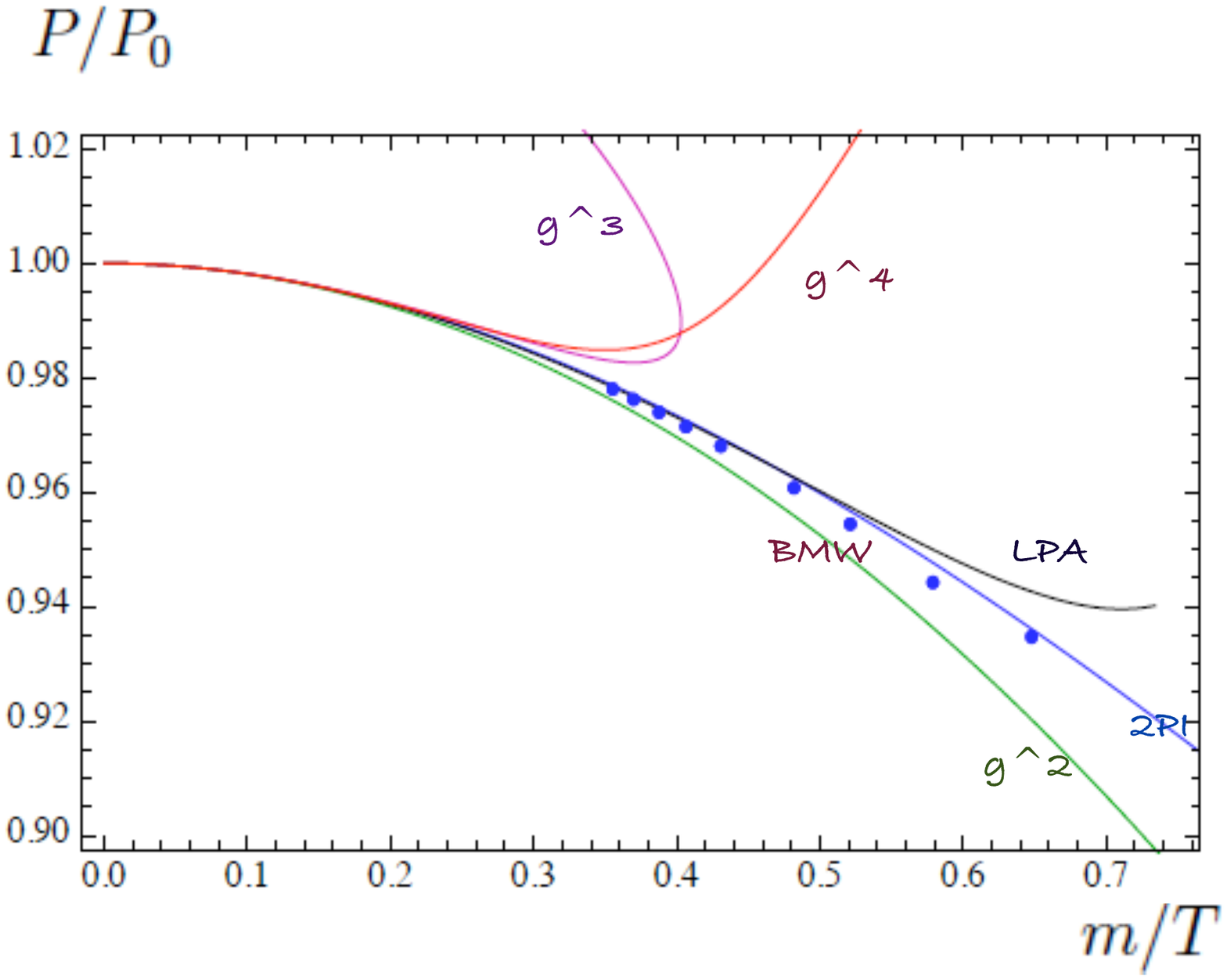}
\vspace{-10mm}
\caption{The pressure as a function of the thermal mass}
\label{fig2}
\end{center}
\end{figure}

\vspace{5mm}

\noindent {\bf Acknowledgements}. I thank Andreas Ipp for his help in preparing Figs. 3 and 4. I am also grateful to the Graduate School of Art and Science of the University of Tokyo for support and hospitality at the time of the write-up of this contribution.


\begin{thebibliography}{99}

\bibitem{RHIC}
  
  RHIC collaborations ``white papers'', Nucl. Phys. A {\bf 757} (2005).
 

\bibitem{Blaizot:2000fc}
  J.~P.~Blaizot, E.~Iancu and A.~Rebhan,
  Phys.\ Rev.\  D {\bf 63} (2001) 065003; 
   Nucl. Phys. A698 (2002) 404.


\bibitem{Endrodi:2007tq}
  G.~Endrodi, Z.~Fodor, S.~D.~Katz and K.~K.~Szabo,
  PoS {\bf LAT2007} (2007) 228.
  
  
\bibitem{Cheng:2008zh}
  M.~Cheng {\it et al.},
  Phys.\ Rev.\  D {\bf 79} (2009) 074505.
  
  
\bibitem{Fodor:2007sy}
  Z.~Fodor,
  PoS {\bf LAT2007} (2007) 011.

 \bibitem{Majumder:2007iu}
  A.~Majumder,
  J.\ Phys.\ G {\bf 34} (2007) S377.
  
  
  \bibitem{Luzum:2008cw}
  M.~Luzum and P.~Romatschke,
  Phys.\ Rev.\  C {\bf 78} (2008) 034915;
  [Erratum-ibid.\  C {\bf 79} (2009) 039903].


\bibitem{Laine:2005ai}
  M.~Laine and Y.~Schroder,
  JHEP {\bf 0503} (2005) 067.



\bibitem{Kajantie:2002wa}
K.~Kajantie, M.~Laine, K.~Rummukainen, and Y.~Schroder,
\newblock {\em Phys. Rev.} D {\bf 67} (2003) 105008.

\bibitem{Blaizot:2003tw}
J.-P. Blaizot, E. Iancu, and A. Rebhan,
\newblock In  {\em Quark-gluon plasma, vol.3*}, R.C. Hwa, editor, World
  Scientific, 2003.

\bibitem{HTL}
  J.~Frenkel and J.~C.~Taylor,
  Nucl.\ Phys.\  B {\bf 334} (1990) 199;
  E.~Braaten and R.~D.~Pisarski,
  Nucl.\ Phys.\  B {\bf 337} (1990) 569.

  \bibitem{Braaten:1995jr}
  E.~Braaten and A.~Nieto,
  Phys.\ Rev.\  D {\bf 53} (1996) 3421. 


\bibitem{Blaizot:2003iq}
  J.~P.~Blaizot, E.~Iancu and A.~Rebhan,
  Phys.\ Rev.\  D {\bf 68} (2003) 025011.


\bibitem{Blaizot:1999ip}
J.~P. Blaizot, E.~ Iancu, and A.~Rebhan,
\newblock {\em Phys. Rev. Lett.} {\bf 83} (1999) 2906--2909.


\bibitem{Boyd:1996bx}
G.~Boyd, J.~Engels, F.~Karsch, E.~Laermann, C.~Legeland, M.~L{\"u}tgemeier, and
  B.~Petersson,
\newblock {\em Nucl. Phys.}, B {\bf 469} (1996) 419--444.

\bibitem{Blaizot:2005wr}
J.-P. Blaizot, A. Ipp, A.~Rebhan, and U.~Reinosa,
\newblock {\em Phys. Rev.}, D {\bf 72} (2005) 125005.



\bibitem{Gubser:2009fc}
  S.~S.~Gubser,
  arXiv:0907.4808 [hep-th].
  
\bibitem{Gubser:1998nz}
S.~S. Gubser, I.~R. Klebanov, and A.~A. Tseytlin,
\newblock {\em Nucl. Phys.}, B {\bf 534} (1998) 202--222.


\bibitem{Blaizot:2006tk}
  J.~P.~Blaizot, E.~Iancu, U.~Kraemmer and A.~Rebhan,
  JHEP {\bf 0706} (2007) 035.


\bibitem{Berges:2000ew}
J.~Berges, N.~Tetradis, and C.~ Wetterich,
\newblock {\em Phys. Rept.} {\bf 363} (2002) 223--386.


\bibitem{Gies:2006wv}
  H.~Gies,
  arXiv:hep-ph/0611146.
  

\bibitem{Blaizot:2006rj}
  J.~P.~Blaizot, A.~Ipp, R.~Mendez-Galain and N.~Wschebor,
  Nucl.\ Phys.\  A {\bf 784} (2007) 376.
  
  \bibitem{Wetterich:1992yh}
C.~Wetterich,
\newblock {\em Phys. Lett.} B {\bf 301} (1993) 90--94.

\bibitem{Blaizot:2005xy}
  J.~P.~Blaizot, R.~Mendez Galain and N.~Wschebor,
  Phys.\ Lett.\  B {\bf 632} (2006) 571.


  
\bibitem{Benitez:2009xg}
  F.~Benitez, J.~P.~Blaizot, H.~Chate, B.~Delamotte, R.~Mendez-Galain and N.~Wschebor,
  Phys.\ Rev.\  E {\bf 80} (2009) 030103.
  

\bibitem{BIW2010}
J.~P.~Blaizot, A.~Ipp and N.~Wschebor, {\em in preparation}. 

  

\bibitem{Gynther:2007bw}
  A.~Gynther, M.~Laine, Y.~Schroder, C.~Torrero and A.~Vuorinen,
  JHEP {\bf 0704}, 094 (2007)



\bibitem{Andersen:2009ct}
  J.~O.~Andersen, L.~Kyllingstad and L.~E.~Leganger,
  JHEP {\bf 0908}, 066 (2009).

  


\end{thebibliography}
\end{document}